# To the Strength First Problem Full Solution: Mechanics of a Necking

S. L. Arsenjev[1]

*Physical-Technical Group*
*Dobrolubova Street, 2, 29, Pavlograd, Dnepropetrovsk region, 51400 Ukraine*

Essentially new approach to analysis of internal forces, arising in cylindrical rod under action of an axial tension force, has allowed to detect the three-dimensional axisymmetric stress state. According to the new conceptual model the axial tension force causes the tangential – hoop and radial – stresses side by side with axial (normal) stress in the rod volume. A completion of Lame's solution of the problem on the stress state in the thick-walled cylinder has allowed to ascertain a mutual direct and reverse connection of the tangential and radial stresses with the axial stress also in the rod under action of an axial tension force. The new approach for the first time has allowed to give the full physically adequate and mathematically sufficiently strict description of a change of the initial cylindrical form of a mild steel rod under action of an axial tension force on all stages of its deforming, including the necking, the fracture process and a view of the fracture surface. The new approach has allowed on the united methodological base to elucidate also a number of the questions, bound with an axisymmetric form of the soap solution film between two rings, with the breaking up of a liquid free jet into drops, with the causes of a buckling of the long tube under action of internal pressure, created in it by the rested and moved fluid in it and others.

**Introduction**
One of the methods of experimental determination of physical and mechanical properties of structural metals and in the first instance of steel is the tension test of a specimen with its working part in the kind of a cylindrical rod. Dimensions of these specimens are standardized, and its diameter is usually 10-20 mm and the relative length of its working part is L/D = 2.5 - 10. The special tension-testing machines are used at the technical and research institutes and at the machine-building plants for a determining of physical and mechanical properties of the new structural metals and also for the control of the strength and plasticity level of the machine, apparatus, structure parts. The tension tests for a determining of static strength are conducted under the limited velocity of the specimen tension. Side by side with strength the plasticity is one of basic characteristics of steel, determining its application. Relative decrease of the specimen cross-section area at a fracture during the tension-test is its plasticity measure. The relative narrowing for most of the structural metals is usually 50 ± 20 % and one is connected with a decrease of the specimen working part diameter in the whole and with a forming of a so called "neck." "Cup and cone" is a name of the fracture surface in the neck zone, on the professional slang. Taking into account the great importance of the plasticity for structural metals and a lack of the satisfactory explanation of a change of the specimen form during the tension test, the author of the given article sets himself as an object the creating of rational conception, which describes the behavior of the steel specimen in the kind of cylindrical rod under action of a tension force physically adequately and mathematically sufficiently strictly. One of basic requirements to the conception is, on the author opinion, a unity of approach to description of a change of the specimen form on all stages of it's deforming from a beginning of the tension test up to the specimen fracture.

---

[1] Phone: (+38 05632) 40596 (home, Rus)
E-mail: usp777@ukr.net; ptglecus@yahoo.com

**Approach**
The approach basis to a solution of the raised problem consists in the following wordy formula:
- a longitudinal tension of a cylindrical rod of isotropic material causes in it the three-dimensional axisymmetric field of internal forces, which are the tangents to an orthogonal family of the cylindrical – dextrorse and sinistrorse – helices; these opposite helices are the action trajectories of the internal forces in the axisymmetric rod under action of a tension force; in the electromagnetism theory the such type lines is called the field lines since Sir Faraday and in the course of Strength of Materials ones are called by the principal stress trajectory;
- on the rod surface these opposite helices form a system of the longitudinal tension forces and of the transversal – hoop – tension forces of equal to them in a quantity;
- the wall boundary layer of the rod has the cross-section area, equal to a half of the rod cross-section area and this layer is under action of the transversal – hoop – tension forces, increasing to the internal boundary of the layer;
- this layer, in the kind of the thick-walled cylindrical shell, compresses the core rod;
- simultaneously both the rod core and compressing of it the rod shell is under action of the same longitudinal tension forces, equal in sum to the tension force, applied to the rod.

**Solution**
In accordance with the approach wordy formula to a solution of the given problem a presentation of the rod as a volume whole in the kind of the rod shell and the rod core allows to use G. Lame's formulas (1833) for an analysis of the transversal – hoop and radial – stresses and of them interaction pressure.
Fig.1 presents diagrams of the forces and stresses for cylindrical rod under action of the axial tension force.
Beforehand it is necessary to take notice, that the hoop stresses are determined by a sum of two transversal forces accordingly to two components – dextrorse and sinistrorse – of the orthogonal family of a field of the cylindrical helices

$$\sigma_t^d + \sigma_t^s = 2\sigma_t ; \sigma_t^d = \sigma_t^s. \tag{1}$$

Every of these two opposite cylindrical helices stipulates an action of the torque in the tensioned rod in a corresponding direction: clockwise and anti-clockwise. These two torque – equal in a quantity and acting in opposite directions – not destroy each other, but ones create the hoop tension stress in the tensioned rod.
Side by side with it an orthogonality of the field lines determines also an equality of the hoop and axial stresses on the rod surface

$$2\sigma_t = 2\sigma_z. \tag{2}$$

Under action of the hoop stresses the rod shell renders a pressure on its core. At the same time the rod shell, in consequence of the great quantity of a ratio of its cross-section radii

$$r_1 = \frac{\sqrt{2}}{2} r_2 \approx 0.707 r_2,$$

can be attributed to the thick-walled cylinders shells under action of internal pressure.
In the considered case Lame's formula for a determining of the hoop stress on external surface of the rod shell has the kind

$$2\sigma_{t(r=r_2)} = p \frac{r_1^2}{r_2^2 - r_1^2} (1 + \frac{r_2^2}{r_2^2}) = 2p = 2\sigma_z, \tag{3}$$

i.e. the pressure quantity on a boundary of the rod shell and the rod core is equal to a quantity of the hoop stress on the rod external surface and simultaneously one is equal to the longitudinal (axial) stress.

The hoop stress in the rod shell on a boundary with the rod core is, accordingly,

$$2\sigma_{t(r=r_1)} = p \frac{r_1^2}{r_2^2 - r_1^2}(1 + \frac{r_2^2}{r_1^2}) = 3p \text{ or } \sigma_{t(r=r_1)} = 1.5\sigma_z. \qquad (4)$$

At that the radial stresses in the rod shell are changed within the limits from a zero on the external surface up to $\sigma_r = p$ on its internal surface.

Simultaneously a well-known condition on a constancy of algebraic sum of the hoop and radial stresses in the thick-walled cylinder under action of the internal pressure

$$\sigma_t + (-\sigma_r) = \text{const}$$

is satisfied all over the cross-section surface of the rod shell and one confirms a constancy of longitudinal stress on this surface.

The hoop and radial stresses in the rod core, taking into consideration it's compactness, has a simple kind

$$\sigma_t = \sigma_r = -p = -\sigma_z$$

all over its cross-section surface.

The adduced expressions for a determining of the stresses in the cylindrical rod under action of the tension force are just both on the elastic and plastic deformation stages while the rod retains a cylindrical form.

At the same time the results of the numerous tension tests of specimens in the kind of the cylindrical rod of the plastic metals and, in particular, of a mild steel show that a form of such specimen is essentially changed during a process of its plastic deforming, although one remains axisymmetric. The general features of the form changes of such specimen at the plastic deforming under action of the tension force are diagrammatically presented in the article [1]. The most typical forms in the process of the specimen irreversible lengthening is the followings:
- the local narrowing of cylindrical part near one of the specimen end;
- a diameter of the specimen cylindrical part is equal to the local narrowing diameter, mentioned on previous stage; the specimen cylindrical part is dilated smoothly up to the initial diameter near both ends of the specimen;
- the specimen cylindrical part takes the shape of a corset;
- the local neck arises in the narrowing part of the corset.

Geometrical analysis of the listed forms allows to ascertain that every of them can be constructed by an adding only two surfaces of rotation to the initial cylindrical surface: the one-hollowed hyperboloid and an evolventoid.

The choice of the one-hollowed hyperboloid is conditioned by two basic reasons. Firstly, this surface can be considered as the cylindrical surface modification if a generatrix of the latter will be split on two crossed inclined straight lines, which ones are symmetrically inclined relatively of the rotation axis and ones connect by itself the circumferences in the cylinder bases. The very visual description of a method of transformation of a cylindrical surface into the one-hollowed hyperboloid surface is adduced by prof. D. Pedoe's in his beautiful book [2]. According to this method a cylinder can be constructed by means of two same wire rings, connected each other by the parallel threads. Then a turn of one of the ring in its plane on some angle, under a condition of the same tension of all threads, transforms the cylinder into the one-hollowed hyperboloid of rotation. In contrast to the cylinder of rotation, possessing only one family of the generatrices, a surface of the hyperboloid contains two families of generatrices. The second family of the generatrices can be formed by a turn of that ring from the initial position in opposite direction that angle. Naturally, a height of such hyperboloid will be less of a height of the initial cylinder proportionally to the incline angle of its generatrices. Secondly, the multitude of the one-hollowed hyperboloids contains a modification, in which the crossing angle of its generatrices, equal $90^0$, and simultaneously its intersection angle with the rotation axis, equal $45^0$. A height of such hyperboloid is equal to the chord length of the circle quarter in its bases. Thus, a height of such hyperboloid is restricted in contrast to a cylinder. The other method of the such hyperboloid construction is that a cube is taken as an initial geometrical figure. If the upper and lower sides of this cube will be circumscribed by circles and the diagonals of one of the cube lateral sides will be rotated about the vertical axis, passing through the centers of the circles, the formed surface of a rotation will be present by itself the one-hollowed hyperboloid with the height, equal to its neck diameter. One of the features of such hyperboloid is that the cross-section area of its neck is equal to a half of its basis area, which approximately corresponds to the relative narrowing of specimens of the a most of the structural metals under a tension test. This special modification of the hyperboloid is called by the regular one-hollowed hyperboloid. In the context of a given problem this hyperboloid modification can be functionally called the orthogonal one-hollowed hyperboloid of rotation. Conformably to a tension of the cylindrical specimen by the axial force, an influence of its form on its stress state is in the following. While the specimen retains the cylindrical form, the orthogonal family of the field lines causes the hoop stresses in its shell, which compress its core by pressure, equal in a quantity to the tension stress. Such action on the specimen core promotes preservation of its continuity during the plastic flow stage up to a reaching of the tension force culmination (ultimate strength). Beginning from this moment, energy of the field of the tension forces reaches a level of the intrastructural energy of the specimen material, preservating its initial form, and then one exceeds the latter. Now the orthogonal family of the internal forces splits a generatrix of the cylindrical surface on two crossing inclined straight lines that are symmetrically intersected with the specimen axis and, in that way, one transforms the specimen cylindrical form into a hyperboloid form, called by a corset. According to experimental data [1] such corset is formed by a turn of one of the base circles of initial cylinder in a beginning on an angle approximately $\pm 32^0$ and then up to approximately $\pm 36^0$. The hyperboloidal corset is simultaneously formed on the whole length of the specimen working part and because of it the lengthening velocity of the specimen becomes commensurable with the action velocity of the tension-testing machine. In a result the tension force, recorded by the machine, is decreased. A decreasing of the specimen cross-section area in the corset neck zone leads to a subsequent increase of the specimen stress state and results in the local narrowing in the kind of the orthogonal hyperboloid. Now, at last, the above mentioned cylindrical helices straighten oneself and coincide with the orthogonal hyperboloid generatrices. In the Strength of Materials course such local narrowing is called by a neck. The neck is formed quickly and accompanied by a decrease of the specimen cross-section area on ~ 50%. Accordingly to such transient and great transversal deformation a lengthening of the specimen is going on again with a velocity commensurable with the action velocity of the tension-testing machine. The machine again records a decrease of the tension force. In the necking process the smooth conjugation of the orthogonal hyperboloid surface with the hyperboloidal corset surface is formed by the rotation surface with its generatrix possessing a variable curvature. The most acceptable curve from the spiral family for such generatrix is a circumference evolvent. While the tensioned specimen retains the cylindrical form the development angle of this evolvent is equal to a zero. In the moment of a forming of the neck, before the specimen fracture, the development angle of this evolvent is $\sim 20^0$ in a point of its conjugation with the surface profile of the orthogonal hyperboloid.

Fig. 2 presents diagrams of a constructing of the longitudinal profile of the tensioned cylindrical rod on the necking stage under action of the axial tension force only. Fig. 3 shows that under action only of the lateral hydrostatic pressure the development angle of the evolvent reaches the maximum quantity $54^0 44`$.

The above mentioned straightening of the cylindrical helices and its coincidence with the orthogonal hyperboloid generatrices results in a ceasing of an acting of the hoop stresses in this part of the specimen and correspondingly in a ceasing of a pressing of the specimen core by the specimen shell. A decrease of the specimen cross-section area on ~50% in the neck corresponds to a doubling of the axial tension stress both in the shell and core parts of the specimen in its minimum cross-section. Under these conditions the specimen shell possesses a possibility for the subsequent plastic deformation owing to a free external surface. In contrast to it such possibility of the specimen core is restricted by a hoop rigidity of the specimen shell. The restriction of possibility of a narrowing of the specimen core in the neck zone under action of the extremal axial tension stress leads to a developing of the three-dimensional tension of the specimen material in this zone. This stress state is in the beginning accompanied by a some additional narrowing of the neck, then by appearance of a porosity of the specimen core in this zone and, at last, by a developing of a transversal crack from the specimen axis in the cross-section, corresponding to the conjugation point of the surface profiles of the orthogonal hyperboloid and the evolvenoid (point C or point E on fig. 2). The subsequent fracture of the specimen shell is going on plastically by the way of a shear owing to the deformation freedom of its external surface and partially of its internal surface. Usually such form of the specimen fracture is called by "cup and cone." One of features of the one-hollowed hyperboloid of rotation is that the cross-section circular form of its neck is lightly transformed into an ellipse. Therefore sometimes [3] the initial transversal crack in the specimen core also takes the ellipse form. The subsequent fracture of the specimen shell is started near the ends of the ellipse major axis and one is completed by the neck-and-split fracture.

In the frames of the above stated approach the mechanics of the local narrowing, spreaded along the specimen working part on the initial stage of its plastic deformation, can be presented by the following way. In an initial state the polycrystalline structure of the mild steel is statistically isotropic and one contains, on the experimental and theoretical research data [4], the rigid intercrystalline lattice of a cementite and the imperfections of the crystals in themselves in the kind of dislocations [5 - 7]. On the initial stage of the specimen plastic deformation an energy of the above described axisymmetric volumetric field of the internal forces, excited by the axial tension force, is sufficiently only for the subsequent fracture of the intercrystalline cementite lattice and only for a deforming of some crystals, possessing a corresponding orientation of the dislocations. As a result of a spreading of the local narrowing along the specimen length its structure is transformed into the statistically orthotropic. In the other words, the initial stage of the specimen plastic deformation is a stage of the volumetric structural adaptation of the specimen material to a structure of the volumetric field of the internal forces, caused by an axial tension force. A surface profile of such local narrowing is, as before, a combination of the hyperboloid with the evolventoid. The results of a solution of the problem, stated in the given article, testify to that the pure uniaxial stressed state is possibly only for the rod with its diameter, equal a zero. In the three-dimensional body the field of the internal forces is also the three-dimensional. At the same time a possibility of a resolution of a three-dimensional field of the internal forces on the orthogonal components is kept in full.

**Discussion of results**

Comparison of the approach and solution of the stated problem with the results of the well-known experimental researches testifies to the following.

Lueders' [8] lines testify to the action possibility of the orthogonal family of the field lines in the kind of the dextrorse and sinistrorse helices in the cylindrical specimen of the mild steel under action of an axial tension force.

E. Bollenrath, V. Hauk and E. Osswald [9], using the roentgenography method, had investigated the residual stresses in the mild steel cylindrical specimen after its preliminary tension to 11%. This research has allowed to ascertain that an area, occupied by the residual axial compressive stresses, has a form of a ring along the external contour of the specimen cross-section and this area is equal to 50% of the specimen cross-section area. The same area is occupied by the residual axial tension stresses in the specimen core.

S. P. Timoshenko [10] cites photos of the short (L/D = 5.4) thin-walled ($\delta$ /D = 0.07) tubes, tensioned up to a fracture. These experiments were conducted by E. A. Davis (Westinghouse Research Laboratories) [11] and ones show that a change of a form of the tubular specimen under action only of a tension force corresponds sufficiently to it in the case of the solid cylindrical specimen.

Among the great volume of information, dedicated to the experimental researches of the mechanical properties of metals by means of the solid cylindrical specimen, P. W. Bridgman's [12] researches are brightly distinguished by a combination of the tension force with the lateral hydrostatic pressure. These researches show that an action of the lateral hydrostatic pressure side by side with the tension force leads to a decrease of the cross-section area of the specimen core in the neck zone. Under sufficiently high lateral pressure the specimen core disappears in this zone and the plastic fracture is going on in the specimen shell. Subsequent increase of the lateral pressure leads to a narrowing of the specimen into a point in the neck zone.

V. I. Feodosjev [13] informs that a plastic deformation and fracture of the cylindrical rod under action of the sufficiently high lateral hydrostatic pressure in the absence of the axial tension force has the character and the shape quite similar to a behavior of that rod under action of the axial tension force only. Using a fig.1 *f*, *g*, it is easy to see that an action of the lateral hydrostatic pressure allows not only to compensate the tension forces between the core and the shell of the rod on the necking stage, but also to ensure a compressing of the rod core by its shell. At that the subsequent tension of the rod will be accompanied, in the beginning, by a narrowing of the rod core and then by a narrowing of the rod shell into a point, similar to the above mentioned Bridgman's experiments.

Thus, the results of the experimental researches corroborate a correctness of the theses of the approach wordy formula to the given problem solution and ones allow to affirm an existence of the direct and reverse connection of the longitudinal (axial) and transversal (hoop and radial) stresses owing to an existence of the orthogonal family of the field lines (the principal stress lines) as a base of the field of internal forces in cylindrical rod. This base exists virtually in the absence of external influence; this base exists really on all stages of elastic and plastic tension. This base exists in the cylindrical rod also under action of the lateral hydrostatic pressure only. In contrast to the explanation [14], in which the necking is bound with imperfection of the rod initial cylindrical form, the solution, given in this article, supposes a geometrically ideal cylindrical form of the rod. As regards the direct and reverse connection of the transversal – hoop and radial – stresses with the axial stress, the above mentioned results of Davis' experiments [11] with the short thin-walled tubes of the mild steel under action of a combination of the axial tension force and the internal hydrostatic pressure testify to the following: the longitudinal residual deformation of the tubular specimen is correspondingly decreased by 32; 39 and 43%, when the designed hoop stresses under action of the internal pressure were some lesser, equal and greater of the designed axial stress. If in Bridgman's experiments the external (lateral) pressure was a cause of the axial tension of the solid cylindrical rod up to its tearing, in Davis' experiments the internal pressure has became an obstacle in the lengthening of the tubular specimen under action of the tension force. Naturally, the steel tube cannot be sufficiently shortened under action only of the internal pressure because of the relatively small quantity of Poisson's ratio of steel. In this case the rubber tube is the best example owing to a quantity of Poisson's ratio of a rubber equal approximately to 0.5.

Thus, in the frames of the approach, stated in the given article, with the taking into account of the cited experimental results the above mentioned well-known condition of a constancy of the axial stress in a cross-section of the thick-walled cylinder under action of the hydrostatic pressure should be described so:

in the case of the internal pressure action

$$\sigma_t + (-\sigma_r) = -\sigma_z, \tag{5}$$

in the case of the external pressure action

$$-\sigma_t = (-\sigma_r) = \sigma_z. \tag{6}$$

As applied to the tensioned cylindrical rod the expressions (5, 6) are same correct for both its shell and core parts and in a whole. The expressions (5, 6) complete a solution of Lame's problem and ones permit to solve the dragged out paradoxal situation, bound with a buckling of the long tube under action of the internal hydrostatic pressure, created by pistons or by the static head of a fluid, flowing through the tube.

J. G. Panovko and I. I. Gubanova [14] declare that such "tube loses the static longitudinal stability, not experiencing of an axial compression force in general."

Feodosjev [13] writes on "a false notion that in the stability question by Euler the internal compression force plays the basic role" and he adds "actually it is not so." At the same time these authors adduce the right solution by Euler, applying formally the axial compression force. It is clear, that such statements are based on the numerous experimental results but not the clear theoretical notion of a given phenomenon. Ascertainment of a connection of the transversal forces with the longitudinal force and a solution of the problem on the internal forces, acting in the cylindrical rod under action of an axial tension force, adduced in the given article, clearly present a mechanics of an origin of the axial internal compression force in the tube under action of the internal pressure both under the static conditions and a fluid flow through the tube.

Just the experiments, considered by authors [13, 14], are an evidence of the direct and reverse connection of the transversal and longitudinal internal forces in these cases.

The modern approach sufficiently corresponds to J. C. Maxwell's conception (1856): "when (the forming energy) reached a certain limit, then the element would be fractured" and also to M. T. Huber's (1904) and A. Foppl's and L. Foppl's (1924) approach, cited by Timoshenko in his textbook [10]. Timoshenko adduces the example of a presentation of the rod stress state under action of a tension force in the kind of a sum of the three-dimensional tension and the pure (simple) shear. The modern approach, stated in this article, presents the stress and strain state field in its concrete form – with taking into account of the body form, of its material and the kind of the load, applied to it.

The results of a research, adduced in the given article, can be applied to a number of the sufficiently far spheres from it.

In particular, the soap water solution film, tensioned between two rings, accepts just the one-hollowed hyperboloid form, ensuring a straightforwardness of the surface tension forces in the absence of a constructive rigidity of the soap solution. Those, who bind a form of such film with a catenoid [15], fail to keep a principle: speak what you know. A drawing of gravitation is misplaced in this case. From two possible surfaces of a rotation – cylindrical and hyperboloidal – the soap solution film, controlled by the intermolecular interaction energy, selects the latter on a principle of the energy minimum for a creating of the form and correspondingly of a minimum of the surface area.

Other example, also bound with a liquid, is a breaking up of the free falling water jet, outflowing out of an orifice (tube), into drops [15]. In this case an increase of the jet fall velocity under action of gravitation causes the longitudinal tension forces in the jet surface and, as a consequence, an increase of the transversal tension forces. In result of it the jet is narrowed similar to the corset of the tensioned steel rod. The smooth narrowing of the jet increases spontaneously the transversal tension forces, leading to the necking in the kind of the one-hollowed hyperboloid. Then a combined action of the longitudinal and transversal tension forces leads to a narrowing of the neck in a point and to a breaking up of the jet. The breaking up is going on very quickly and one is accompanied by a sharp rebuilding of the hyperboloid form into two semispheres. This transient process, in one's turn, is accompanied by a strong wave impulse, acting into the jet upper and lower parts. Under action of this impulse the longitudinal and transversal forces of the surface tension tear almost instantly the jet lower part into drops. Thus the described very upper break of the water jet is a generator of its breaking up into drops.

Next example, also bound with the hyperboloid form, is form of a bell from the great at the church belfry to the ship bell, hand bell. Sir Rayleigh writes in his fundamental work [16] "as regards a form, accepted for the church bells, one is not received the satisfactory explanation till now." The Great Soviet Encyclopedia [17] contains the following thoughtful description: "the bell has a form of the hollow pear without its lower part." Prof. Pedoe notices in above mentioned his book [2] that the neck cross-section of the one-hollowed hyperboloid, constructed by means of two wire rings and the threads between its, has the ellipse form. The author of the given article has availed oneself of a diversity of the modern form of the polyethylene bottles and has purchased a bottle, contained the corset with 7.6 cm neck diameter, and the 9 cm diameter of its bottoms and a distance between the bottoms equal 21 cm. Both bottoms of this bottle have the kind similar to a sphere and ones have a sufficient rigidity of its form. At the same time hoop rigidity of the corset turned out by extremely low. The slightest drawing out of air out of the bottle transforms right away a round cross-section of the corset neck into ellipse. The slightest supercharging of the bottle transforms the ellipse into a circle. The corset bases (the bottle bottoms) in both these cases keep the round form of its cross-section. A tapping by a finger tip along the corset from one bottom to the other has allowed to detect on both sides from the corset neck a singly belt, where the tapping, not causes the characteristic sounding. The tapping outside these silence belts in both the corset neck zone and the corset bases causes the characteristic low-frequency sounding with its attenuation during approximately 7 seconds. These silence belts is located in the junction of the hyperboloid neck concave profile with the convex profile beyond the neck. The same tapping on a surface of a cylindrical polyethylene bottle not causes the pure tone and lasts no more 1.5 – 2 seconds. In a next experiment with the corset bottle the tapping was accompanied by a supercharging of the bottle. The experiment has shown the strong direct connection of the sound frequency with the supercharge pressure at the same duration of the sound attenuation. Then the corset bottle was shut off by the threaded lid and placed in the warm air stream. Under this condition the tapping has allowed to ascertain that the heating in limits from ~ $20^0$ up to ~ $40^0$ C leads to a rising of the sound tone practically on an octave. An observation of the corset surface and the listening of it in a process of the sound oscillations have allowed to ascertain that just the corset neck zone is a base source of the sound radiation. For all this the corset neck cross-section accepts an ellipse form with a changing of its major and minor axes at every cycle of the oscillation and correspondingly to it the directional diagram of the sound radiation has four petals, and one of them coincides with a place of the tapping. So, the basic constructive features of the bell can be presented in the kind of two versions. Accordingly to the first of them these features are the followings:
- a form of middle surface on the bell lateral shell is the one-hollowed hyperboloid part, contained the neck part and adjoining to it from below the diverging part;
- an upper part of the bell lateral shell is restricted by a roof in the kind of a sphere segment with the tangent angle to its base plane ~ $35^0$; the roof contains in its centre the bell hanger brackets;
- a lower part of the bell lateral shell is broadened to a diameter equal 0.75 – 0.8 of a height of the bell lateral shell and one has along the lower edge a thickening – the percussive belt;
- in the whole the bell lateral shell has approximately the same relative thickness.

In this case the belt shell not contains the silence belts.
Accordingly to the second version the lower part of the bell lateral shell presents by itself the convexo-convex shell, forming the silence belt with the bell hyperboloidal neck.
Fig. 4 shows diagrammatically the mentioned versions of the bell construction.
Owing to the silence belt, placed between the bell neck and the bell lower part, these two parts of the bell lateral shell oscillate in opposite phases. In both described versions frequency of the sound oscillations, radiated by the bell, is inversely proportional to its neck diameter. The described structure of the bell form allows to give the concrete recommendations for an assuring of the radiated sound tone, but it oversteps the frames of the given article.
Although a detection of the silence belts allows to explain why the initial transversal crack in the core of the tensioned cylindrical specimen, as stated above, arises just in the plane, passing through the junction of the hyperboloid and evolventoid surfaces. The fact is that the deformation and plastic flow of the specimen are accompanied by the longitudinal and transversal oscillations of a sufficiently high magnitude. A change of a curvature of the longitudinal profile from the convex in the evolventoid zone to the concave in the hyperboloid zone leads to that the transversal (radial) oscillations are going on in the opposite phases in these zones. In a result the above plate is a plate of an action of the radial cyclic shear stress, promoting a shaking of the specimen material structure. In this case the above silence belt renders the negative influence on the specimen strength.
Reverting to the base object of the given article, it is also necessary to note, that the necking is the typical phenomenon for such metals as the mild steel. At the same time the features of the metal internal structure lead not only to the different modifications of the necking, but to a fracture without the necking. For example, the cylindrical specimen of the grey iron under action of an axial tension force is fractured without the visible residual deformation (the square-break fracture). Such distinction of the specimen behavior is explained by that the ultimate strength of the grey iron is in 2 – 5 times less than its compression strength. Therefore, in contrast to the steel specimen the grey iron specimen fracture is going on by means of a forming of the transversal hoop crack on its external surface and its subsequent fast developing to the specimen axis.

Suppose so, with the taking into account of the above mentioned Bridgman's experiments, that the grey iron specimen will be fractured with the necking under action of a combination of the sufficiently high hydrostatic pressure and the comparatively not great axial compressive force.

Next question, affected in the end of the Solution section of the given article – on pure uniaxial stressed state – has the great methodological significance and one requires the more detailed explanation. The fact is that the traditional course of Strength of Materials from the very outset presents the rod tension as a "simple tension." The determination of a stressed state in it is restricted by a division of the tension force by the rod cross-section area. This "simple tension" is then presented as the uniaxial stressed state. Further the traditional approach envisages a study of the planar stressed state and then the triaxial (3-D) stressed state as a complex stressed state. In that way the traditional teaching is formally constructed rightly - from a simple to a complex. At the same time the results of the tension tests of the thin wires of the mild metals (with the wire diameter, compared with the grain diameter of the wire metals) show that in these cases the fracture is going on without the necking (see, for example, I. J. Dechtjar`s article in the above mentioned the Collected articles [3]). This feature in a comparison with the results of the tension tests of the standard specimens of the mild steel testifies to that the rod with a diameter 10 – 20 mm is presented in the kind of a bunch of the thin wires with the total cross-section area, equal the solid rod cross-section area. In a result those, who had good studied the Strength of Materials, ones are astonished by the necking and those, who had weakly studied this course, ones do not think about it. This paradoxal situation exists more than 100 years.

In this connection it is appropriate to appeal to prof. L. R. G. Treloar`s book [18], in which the author adduces an example of a tension of the usual polyethylene film strip 10 cm long and 1 cm width. In this experiment the tension is accompanied by shoulder effect in the kind of the local narrowing of the strip, which envelops gradually all its tensioned part similar to the local narrowing, enveloped the all length of the mild steel rod, tensioned on the initial stage of its plastic deformation. But, in contrast to the steel rod, the polyethylene strip fracture is going on without the necking, i. e. practically brittle. Such fracture of the very plastic material is stipulated by the following. On the initial stage of the plastic deformation the polyethylene molecular chains straighten oneself and ones are stretched along the tension direction, losing a part of its transversal ties. In a result of it the entire material is transformed practically into a bunch of the microfibres, the every of which and all ones together are fractured without the necking. Such entire material is more similar to the wire rope than the solid rod.

This is what, in brief, follows from the problem solution on the "simple tension" and on the necking.

In the discussion conclusion it is necessary to note that the successive solution of the stated problem is the next in turn example of a using of the Physical Ensemble method, developed by author of the given article. An effectiveness of the method for a physically adequate and mathematically sufficiently strict construction of the models of the some phenomena in a field of the fluid dynamics was showed in previous article [19], written by the author separately, and in articles [20 – 28], written by the author with his collaborators.

**Final remarks**
The approach, the problem solution and the considered examples, stated in given article, create the necessary prerequisites for the fruitful solution of the problems and questions, bound with an analysis of the strained and stressed state of the solid and liquid on the new united physically sensible methodological base.

**Acknowledgements**
Author wants to express his profound respect to the outstanding precursors, supposing that the modern results, filling the mechanics separate gaps, are stipulated by the natural development of the great heritage. Author wants to express his deep gratitude to his son Alexey for typing of this article text and carrying out of this article graphical part. Author wants also to express his deep gratitude to his daughter Catherina for her caring for her father. Author dedicates this article to the unforgettable memory of Nina Stepanovna Nikitenko the talented specialist, lecturer on Strength of Materials at Polytechnic Institute of Odessa, 1958 – 1963 and 1965.

_______________________________

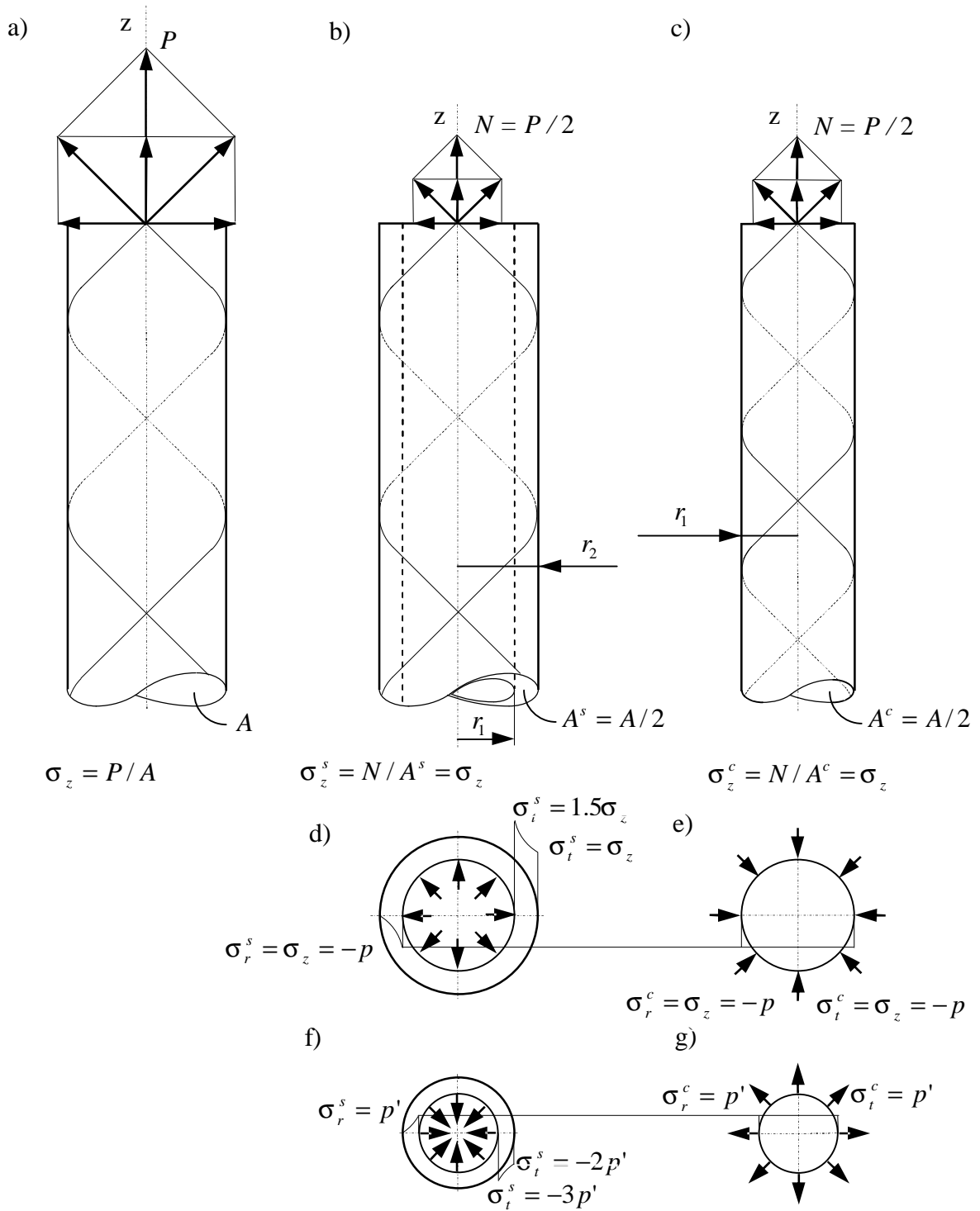

Fig.1. Diagrams of the forces and stresses for the cylindrical rod under action of the axial tension force: a,b,c) diagrams of the forces and the field lines, acting in the rod and its part – shell (superscript *s*) and core (superscript *c*) – correspondingly; d,e) diagrams of the transversal – hoop and radial – stresses in the rod shell and in the rod core, correspondingly, and its mutual pressure in the initial stage of the rod plastic deformation; f,g) diagrams of the transversal – hoop and radial – stresses in the rod shell and in the rod core, correspondingly, and its mutual pressure in the neck zone

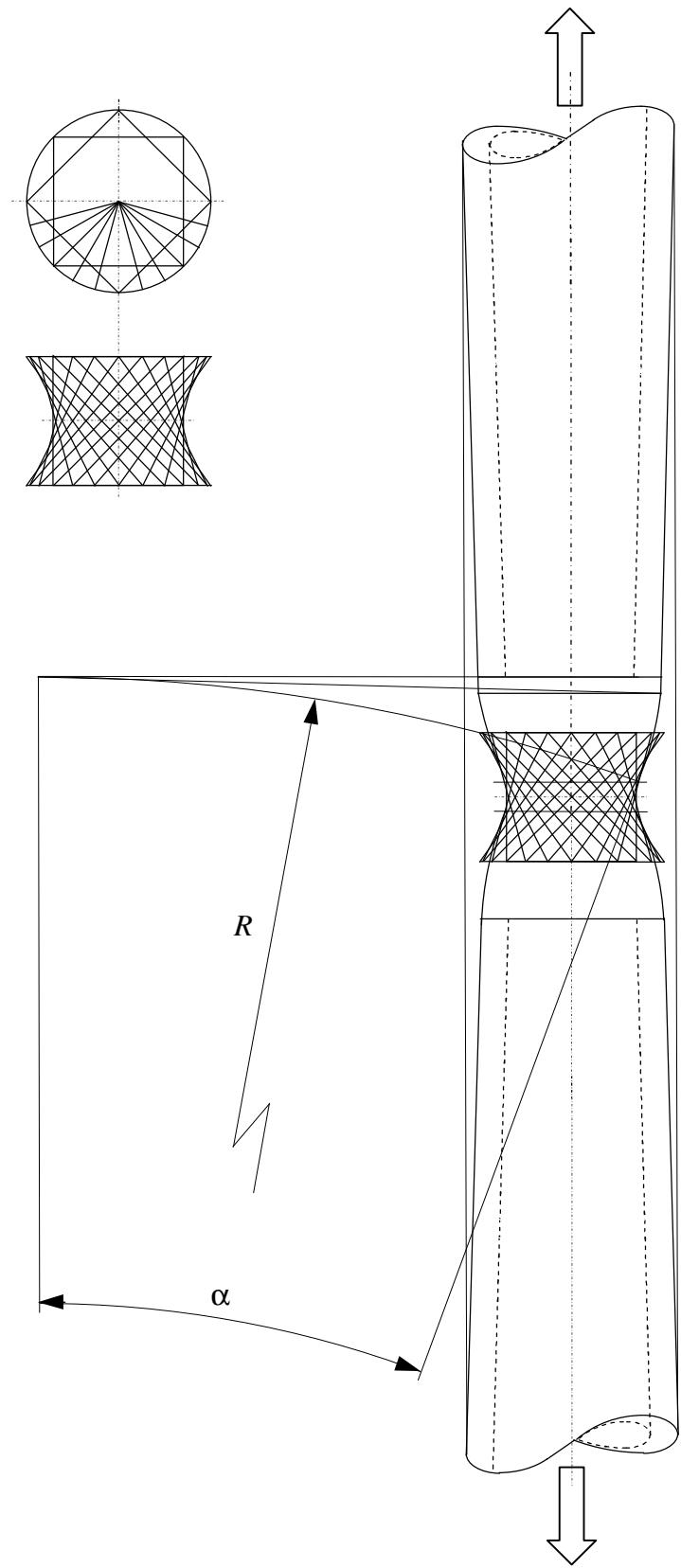

Fig. 2. Diagrams of a constructing of the one-hollowed hyperboloid and of the longitudinal profile of the tensioned cylindrical rod on the necking stage under action of the axial tension force only

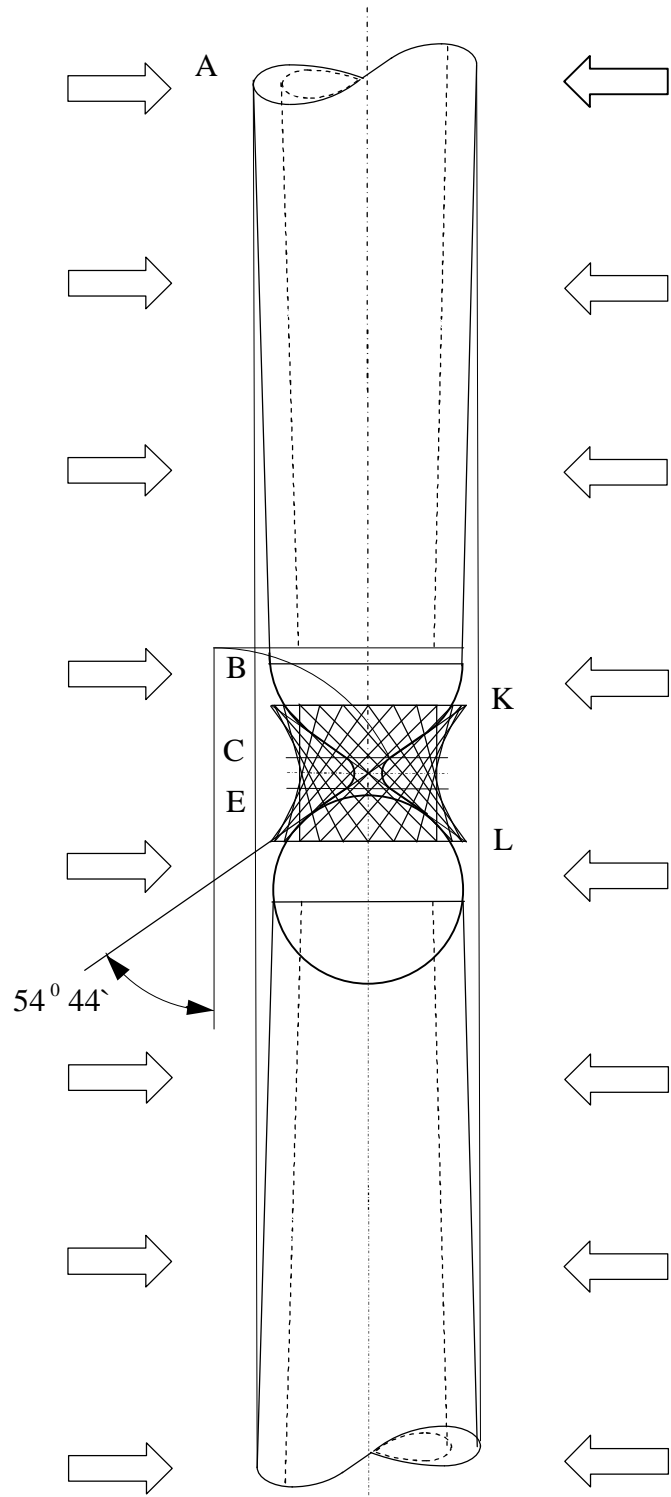

Fig. 3. Diagrams of a constructing of the longitudinal profile of the tensioned cylindrical rod on the necking stage under action only of the lateral hydrostatic pressure: the development angle of the evolvent reaches the maximum quantity 54 $^0$ 44`; AB – the hyperboloidal corset; BC – the evolventiodal belt; CE – the orthogonal hyperboloid belt; KL – the orthogonal hyperboloid in the hyperboloidal corset

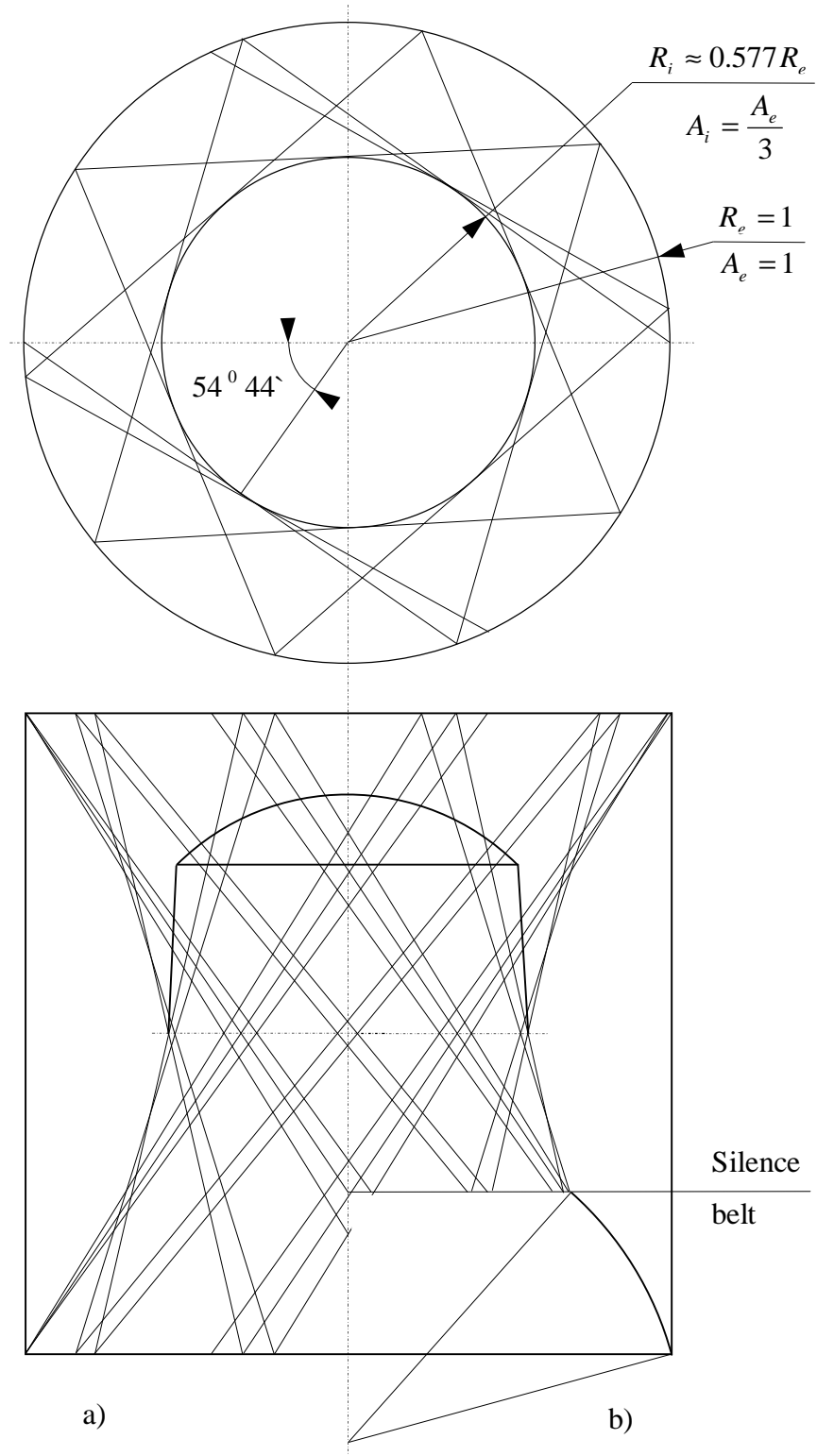

Fig. 4. The construction diagram of the middle surface profile of the bell (the Tsar-bell in the Moscow Kremlin): a) combination of hyperboloid and cone; b) combination of hyperboloid, cone and torus